\documentclass{aastex63}

\usepackage[utf8x]{inputenc}
\usepackage{amssymb}
\usepackage{amsmath}
\usepackage{natbib}
\usepackage{hyperref}

\received{\today}
\revised{\today}
\accepted{\today}
\submitjournal{ApJL}

\shorttitle{Coronal Heating Rate in the Slow Solar Wind}
\shortauthors{Telloni et al.}

\begin{document}

\title{Coronal Heating Rate in the Slow Solar Wind}

\correspondingauthor{Daniele Telloni}
\email{daniele.telloni@inaf.it}

\author[0000-0002-6710-8142]{Daniele Telloni}
\affil{National Institute for Astrophysics, Astrophysical Observatory of Torino, Via Osservatorio 20, I-10025 Pino Torinese, Italy}
\author[0000-0001-9921-1198]{Marco Romoli}
\affil{University of Florence, Department of Physics and Astronomy, Via Giovanni Sansone 1, I-50019 Sesto Fiorentino, Italy}
\author[0000-0002-2381-3106]{Marco Velli}
\affil{Earth, Planetary, and Space Sciences, University of California, Los Angeles, CA 90095, USA}
\author[0000-0002-4642-6192]{Gary P. Zank}
\affil{Center for Space Plasma and Aeronomic Research, University of Alabama in Huntsville, Huntsville, AL 35805, USA}
\affil{Department of Space Science, University of Alabama in Huntsville, Huntsville, AL 35805, USA}
\author[0000-0003-1549-5256]{Laxman Adhikari}
\affil{Center for Space Plasma and Aeronomic Research, University of Alabama in Huntsville, Huntsville, AL 35805, USA}
\author[0000-0003-1759-4354]{Cooper Downs}
\affil{Predictive Science Inc., San Diego, CA 92121, USA}
\author[0000-0002-8734-808X]{Aleksandr Burtovoi}
\affil{National Institute for Astrophysics, Astrophysical Observatory of Arcetri, Largo Enrico Fermi 5, I-50125 Firenze, Italy}
\author[0000-0002-1017-7163]{Roberto Susino}
\affil{National Institute for Astrophysics, Astrophysical Observatory of Torino, Via Osservatorio 20, I-10025 Pino Torinese, Italy}
\author[0000-0003-3517-8688]{Daniele Spadaro}
\affil{National Institute for Astrophysics, Astrophysical Observatory of Catania, Via Santa Sofia 78, I-95123 Catania, Italy}
\author[0000-0002-4299-0490]{Lingling Zhao}
\affil{Center for Space Plasma and Aeronomic Research, University of Alabama in Huntsville, Huntsville, AL 35805, USA}
\author[0000-0002-0016-7594]{Alessandro Liberatore}
\affil{Jet Propulsion Laboratory, California Institute of Technology, Pasadena, CA 91109, USA}
\author[0000-0002-2582-7085]{Chen Shi}
\affil{Earth, Planetary, and Space Sciences, University of California, Los Angeles, CA 90095, USA}
\author[0000-0003-2426-2112]{Yara De Leo}
\affil{Max Planck Institute for Solar System Research, Justus-von-Liebig-Weg 3, D-37077 G\"ottingen, Germany}
\affil{University of Catania, Department of Physics and Astronomy, Via Santa Sofia 64, I-95123 Catania, Italy}
\author[0000-0001-8235-2242]{Lucia Abbo}
\affil{National Institute for Astrophysics, Astrophysical Observatory of Torino, Via Osservatorio 20, I-10025 Pino Torinese, Italy}
\author[0000-0001-9014-614X]{Federica Frassati}
\affil{National Institute for Astrophysics, Astrophysical Observatory of Torino, Via Osservatorio 20, I-10025 Pino Torinese, Italy}
\author[0000-0002-0764-7929]{Giovanna Jerse}
\affil{National Institute for Astrophysics, Astronomical Observatory of Trieste, Localit\`a Basovizza 302, I-34149 Trieste, Italy}
\author[0000-0001-8244-9749]{Federico Landini}
\affil{National Institute for Astrophysics, Astrophysical Observatory of Torino, Via Osservatorio 20, I-10025 Pino Torinese, Italy}
\author[0000-0002-9459-3841]{Gianalfredo Nicolini}
\affil{National Institute for Astrophysics, Astrophysical Observatory of Torino, Via Osservatorio 20, I-10025 Pino Torinese, Italy}
\author[0000-0002-3789-2482]{Maurizio Pancrazzi}
\affil{National Institute for Astrophysics, Astrophysical Observatory of Torino, Via Osservatorio 20, I-10025 Pino Torinese, Italy}
\author[0000-0002-2433-8706]{Giuliana Russano}
\affil{National Institute for Astrophysics, Astronomical Observatory of Capodimonte, Salita Moiariello 16, I-80131 Napoli, Italy}
\author[0000-0002-5163-5837]{Clementina Sasso}
\affil{National Institute for Astrophysics, Astronomical Observatory of Capodimonte, Salita Moiariello 16, I-80131 Napoli, Italy}
\author[0000-0003-1962-9741]{Vincenzo Andretta}
\affil{National Institute for Astrophysics, Astronomical Observatory of Capodimonte, Salita Moiariello 16, I-80131 Napoli, Italy}
\author[0000-0001-6273-8738]{Vania Da Deppo}
\affil{National Research Council, Institute for Photonics and Nanotechnologies, Via Trasea 7, I-35131 Padova, Italy}
\author[0000-0002-2789-816X]{Silvano Fineschi}
\affil{National Institute for Astrophysics, Astrophysical Observatory of Torino, Via Osservatorio 20, I-10025 Pino Torinese, Italy}
\author[0000-0002-5467-6386]{Catia Grimani}
\affil{University of Urbino Carlo Bo, Department of Pure and Applied Sciences, Via Santa Chiara 27, I-61029 Urbino, Italy}
\affil{National Institute for Nuclear Physics, Section in Florence, Via Bruno Rossi 1, I-50019 Sesto Fiorentino, Italy}
\author[0000-0002-5778-2600]{Petr Heinzel}
\affil{Czech Academy of Sciences, Astronomical Institute, Fri\v{c}ova 298, CZ-25165 Ond\v{r}ejov, Czechia}
\affil{University of Wroc\l{}aw, Centre of Scientific Excellence -- Solar and Stellar Activity, ul. Kopernika 11, PL-51-622 Wroc\l{}aw, Poland}
\author[0000-0001-9670-2063]{John D. Moses}
\affil{National Aeronautics and Space Administration, Headquarters, Washington, DC 20546, USA}
\author[0000-0003-2007-3138]{Giampiero Naletto}
\affil{University of Padua, Department of Physics and Astronomy, Via Francesco Marzolo 8, I-35131 Padova, Italy}
\author[0000-0002-5365-7546]{Marco Stangalini}
\affil{Italian Space Agency, Via del Politecnico snc, I-00133 Roma, Italy}
\author[0000-0001-7298-2320]{Luca Teriaca}
\affil{Max Planck Institute for Solar System Research, Justus-von-Liebig-Weg 3, D-37077 G\"ottingen, Germany}
\author[0000-0002-7585-8605]{Michela Uslenghi}
\affil{National Institute for Astrophysics, Institute of Space Astrophysics and Cosmic Physics of Milan, Via Alfonso Corti 12, I-20133 Milano, Italy}
\author[0000-0002-6505-4478]{Arkadiusz Berlicki}
\affil{University of Wroc\l aw, Astronomical Institute, Kopernika 11, PL-51622 Wroc\l aw, Poland}
\author[0000-0002-2152-0115]{Roberto Bruno}
\affil{National Institute for Astrophysics, Institute for Space Astrophysics and Planetology, Via del Fosso del Cavaliere 100, I-00133 Roma, Italy}
\author[0000-0003-0520-2528]{Gerardo Capobianco}
\affil{National Institute for Astrophysics, Astrophysical Observatory of Torino, Via Osservatorio 20, I-10025 Pino Torinese, Italy}
\author[0000-0002-8430-8218]{Giuseppe E. Capuano}
\affil{National Institute for Astrophysics, Astrophysical Observatory of Catania, Via Santa Sofia 78, I-95123 Catania, Italy}
\affil{University of Catania, Department of Physics and Astronomy, Via Santa Sofia 64, I-95123 Catania, Italy}
\author[0000-0001-8783-0047]{Chiara Casini}
\affil{National Research Council, Institute for Photonics and Nanotechnologies, Via Trasea 7, I-35131 Padova, Italy}
\affil{Centre of Studies and Activities for Space ``Giuseppe Colombo'', Via Venezia 15, I-35131 Padova, Italy}
\author[0000-0002-9716-3820]{Marta Casti}
\affil{The Catholic University of America at the National Aeronautics and Space Administration, Goddard Space Flight Center, Greenbelt, MD 20771, USA}
\author[0000-0002-3379-2142]{Paolo Chioetto}
\affil{National Research Council, Institute for Photonics and Nanotechnologies, Via Trasea 7, I-35131 Padova, Italy}
\affil{Centre of Studies and Activities for Space ``Giuseppe Colombo'', Via Venezia 15, I-35131 Padova, Italy}
\author[0000-0003-0378-9249]{Alain J. Corso}
\affil{National Research Council, Institute for Photonics and Nanotechnologies, Via Trasea 7, I-35131 Padova, Italy}
\author[0000-0003-2647-117X]{Raffaella D'Amicis}
\affil{National Institute for Astrophysics, Institute for Space Astrophysics and Planetology, Via del Fosso del Cavaliere 100, I-00133 Roma, Italy}
\author[0000-0002-2464-1369]{Michele Fabi}
\affil{University of Urbino Carlo Bo, Department of Pure and Applied Sciences, Via Santa Chiara 27, I-61029 Urbino, Italy}
\affil{National Institute for Nuclear Physics, Section in Florence, Via Bruno Rossi 1, I-50019 Sesto Fiorentino, Italy}
\author[0000-0001-5528-1995]{Fabio Frassetto}
\affil{National Research Council, Institute for Photonics and Nanotechnologies, Via Trasea 7, I-35131 Padova, Italy}
\author[0000-0002-4453-1597]{Marina Giarrusso}
\affil{University of Florence, Department of Physics and Astronomy, Via Giovanni Sansone 1, I-50019 Sesto Fiorentino, Italy}
\author[0000-0002-3468-8566]{Silvio Giordano}
\affil{National Institute for Astrophysics, Astrophysical Observatory of Torino, Via Osservatorio 20, I-10025 Pino Torinese, Italy}
\author[0000-0002-1837-2262]{Salvo L. Guglielmino}
\affil{National Institute for Astrophysics, Astrophysical Observatory of Catania, Via Santa Sofia 78, I-95123 Catania, Italy}
\author[0000-0002-0901-0251]{Enrico Magli}
\affil{Politecnico of Turin, Department of Electronics and Telecommunications, Corso Duca degli Abruzzi 24, I-10129 Torino, Italy}
\author[0000-0002-2656-1557]{Giuseppe Massone}
\affil{National Institute for Astrophysics, Astrophysical Observatory of Torino, Via Osservatorio 20, I-10025 Pino Torinese, Italy}
\author[0000-0002-5422-1963]{Mauro Messerotti}
\affil{National Institute for Astrophysics, Astronomical Observatory of Trieste, Localit\`a Basovizza 302, I-34149 Trieste, Italy}
\affil{University of Trieste, Department of Physics, Via Alfonso Valerio 2, I-34127 Trieste, Italy}
\author[0000-0003-2566-2820]{Giuseppe Nistic\`o}
\affil{University of Calabria, Department of Physics, Ponte Pietro Bucci Cubo 31C, I-87036 Rende, Italy}
\author[0000-0002-1383-6750]{Maria G. Pelizzo}
\affil{University of Padua, Department of Information Engineering, Via Giovanni Gradenigo 6, I-35131 Padova, Italy}
\author[0000-0002-1820-4824]{Fabio Reale}
\affil{University of Palermo, Department of Physics and Chemistry - Emilio Segr\`e, Piazza del Parlamento 1, I-90134 Palermo, Italy}
\affil{National Institute for Astrophysics, Astronomical Observatory of Palermo, Piazza del Parlamento 1, I-90134 Palermo, Italy}
\author[0000-0001-7066-6674]{Paolo Romano}
\affil{National Institute for Astrophysics, Astrophysical Observatory of Catania, Via Santa Sofia 78, I-95123 Catania, Italy}
\author[0000-0001-6060-9078]{Udo Sch\"uhle}
\affil{Max Planck Institute for Solar System Research, Justus-von-Liebig-Weg 3, D-37077 G\"ottingen, Germany}
\author[0000-0002-3418-8449]{Sami K. Solanki}
\affil{Max Planck Institute for Solar System Research, Justus-von-Liebig-Weg 3, D-37077 G\"ottingen, Germany}
\author[0000-0002-6280-806X]{Thomas Straus}
\affil{National Institute for Astrophysics, Astronomical Observatory of Capodimonte, Salita Moiariello 16, I-80131 Napoli, Italy}
\author[0000-0002-5152-0482]{Rita Ventura}
\affil{National Institute for Astrophysics, Astrophysical Observatory of Catania, Via Santa Sofia 78, I-95123 Catania, Italy}
\author[0000-0002-4997-1460]{Cosimo A. Volpicelli}
\affil{National Institute for Astrophysics, Astrophysical Observatory of Torino, Via Osservatorio 20, I-10025 Pino Torinese, Italy}
\author[0000-0002-4184-2031]{Luca Zangrilli}
\affil{National Institute for Astrophysics, Astrophysical Observatory of Torino, Via Osservatorio 20, I-10025 Pino Torinese, Italy}
\author[0000-0002-9207-2647]{Gaetano Zimbardo}
\affil{University of Calabria, Department of Physics, Ponte Pietro Bucci Cubo 31C, I-87036 Rende, Italy}
\author[0000-0003-0290-3193]{Paola Zuppella}
\affil{National Research Council, Institute for Photonics and Nanotechnologies, Via Trasea 7, I-35131 Padova, Italy}
\author[0000-0002-1989-3596]{Stuart D. Bale}
\affil{Space Sciences Laboratory, University of California, Berkeley, CA 94720, USA}
\affil{Physics Department, University of California, Berkeley, CA 94720, USA}
\author[0000-0002-7077-930X]{Justin C. Kasper}
\affil{BWX Technologies, Inc., Washington, DC 20002, USA}
\affil{Climate and Space Sciences and Engineering, University of Michigan, Ann Arbor, MI 48109, USA}

\begin{abstract}
This Letter reports the first observational estimate of the heating rate in the slowly expanding solar corona. The analysis exploits the simultaneous remote and local observations of the same coronal plasma volume with the Solar Orbiter/Metis and the Parker Solar Probe instruments, respectively, and relies on the basic solar wind magnetohydrodynamic equations. As expected, energy losses are a minor fraction of the solar wind energy flux, since most of the energy dissipation that feeds the heating and acceleration of the coronal flow occurs much closer to the Sun than the heights probed in the present study, which range from $6.3$ to $13.3$ R$_{\odot}$. The energy deposited to the supersonic wind is then used to explain the observed slight residual wind acceleration and to maintain the plasma in a non-adiabatic state. As derived in the Wentzel-Kramers-Brillouin limit, the present energy transfer rate estimates provide a lower limit, which can be very useful in refining the turbulence-based modeling of coronal heating and subsequent solar wind acceleration.
\end{abstract}


\section{Introduction}
\label{sec:introduction}
Central to the heating and subsequent acceleration of the coronal plasma is the identification of the physical mechanisms responsible for transporting the energy available in the photospheric motions to the corona, where it is dissipated by raising the temperature to a million degrees, sufficient for the plasma to overcome the Sun's gravity and thus expand into the heliosphere, forming the solar wind \citep{1958ApJ...128..664P}. It is worth noting that this thermal driving can only account for the acceleration of the slowest coronal flows, while for the higher speed streams to be accelerated, an additional source of energy is required, whose dissipation results in further heating and thus transition to a faster wind regime. The non-resonant dissipation of low-frequency MagnetoHydroDynamic (MHD) turbulent fluctuations is to date the most widely accepted heating mechanism, whether for fast or slow wind.

However, a key element for any heating model, namely the estimate of the coronal energy transfer rate and its radial dependence, is still missing. This is basically because coronal measurements of magnetic field and non-thermal plasma motions (related to turbulence), i.e., Alfv\'en wave energy, are lacking (or at best inconclusive). In some fluid models \citep[e.g.,][]{1970ARA&A...8...31H,1995A&A...303L..45M}, an ad hoc heating function, which decays exponentially with height over a prescribed dissipation length scale, is assumed to account for the energy deposition per unit volume. Other approaches resort to the use of numerical simulations in order to derive the coronal heating rate profile \citep[e.g.,][]{2002ApJ...575..571D}.

Local single-point measurements of the turbulent energy cascade rate, which is directly related to the solar wind heating rate, are widely available at various heliospheric heights. Although a detailed discussion of the turbulent heating rate estimates in the solar wind is beyond the scope of this article \citep[the interested reader is referred to the comprehensive review by][]{2023PhR..1006....1M}, the following works are worth mentioning. \citet{2007PhRvL..99k5001S} first applied the generalized form of the Yaglom law for MHD turbulence \citep[as derived by][]{1998GeoRL..25..273P} to Ulysses data \citep{1992A&AS...92..207W} to derive the energy flux of the turbulent cascade in high-speed solar streams sampled at $3-4$ au. \citet{2022ApJ...928L..15Z} extended this approach to the first sub-Alfv\'enic interval observed with Parker Solar Probe \citep[PSP;][]{2016SSRv..204....7F} at $0.09$ au, using both incompressible and compressible formalisms for the equivalent Yaglom law \citep{2019PhRvL.123x5101A}, and finding a higher average energy cascade rate than in the surrounding super-Alfv\'enic regions. Adopting the turbulence cascade model developed by \citet{1988JGR....93....7T} and exploiting PSP measurements, \citet{2020ApJ...904L...8W} derived a formula for relating the radial scaling of the low-frequency spectral break to the energy supply rate in the slow solar wind, from $0.17$ to $0.7$ au. These authors found that at distances greater than $0.25$ au from the Sun, injected and dissipated energies are of the same order, suggesting that the slow solar wind expands almost adiabatically. Deriving straightforward expressions for the turbulent heating rate, \citet{2022ApJ...938..120A} reported, from both observational and theoretical perspectives, on the radial evolution of the energy dissipation rate from $0.17$ to $0.83$ au, using PSP and Solar Orbiter \citep[SO;][]{2020A&A...642A...1M} data. It is true that some authors have exploited remote observations of the solar corona to probe the turbulent properties of coronal flows and thus derive constraints or upper limits on the rate of energy deposition in the corona \citep{2009ApJ...707.1668C,2020ApJ...900..105C,2021ApJ...914..137S}. Yet, all these studies make extensive use of unknown parameter assumptions and are either highly model-dependent or based on numerical simulations. Although extremely important, these works do not provide direct, empirical observations of the coronal heating rate as in the present study.

Quadratures between two spacecraft have already proved useful in linking local turbulent properties with the fluid parameters of remotely observed coronal flows \citep{2021ApJ...920L..14T,2022ApJ...935..112T}. However, the special orbital configuration between PSP and SO spacecraft that occurred on June $01$, $2022$ offers an almost unique opportunity. Indeed, during this quadrature, PSP entered the portion of the Plane of the Sky (POS) imaged by the Metis coronagraph \citep{2020ExA....49..239F,2020A&A...642A..10A} on board SO. As described in the following, this allows the first ever observational estimate of the energy deposition rate in the solar corona, from $6.3$ to $13.3$ R$_{\odot}$, without having to invoke any model or ad hoc parameter assumptions. Indeed, the simultaneous remote and in-situ measurements of the same coronal flow, associated with solving the basic equations for a steady-state, radial plasma flow, allow an empirical low-speed solar wind model to be derived. In addition, in the Wentzel-Kramers-Brillouin (WKB) theory approximation, the energy of the outward Alfv\'en wave flux and, in turn, the energy transfer rate, can be extended from the PSP location to the whole range of coronal heights observed with Metis. This is the aim of the present Letter, whose plan is as follows: Metis/PSP data and equation description (\S{} \ref{sec:data_equation}), derivation of the empirical wind and turbulent model along with relevant discussion (\S{} \ref{sec:empirical_model}).

\section{SO/Metis -- PSP data and MHD equations}
\label{sec:data_equation}
On June $01$, $2022$, at $22$:$40$ UT, when SO was at $0.936$ au, PSP was in quadrature while orbiting at $13.3$ R$_{\odot}$. A SO spacecraft roll of $45^{\circ}$ along with an off-pointing of $1$ R$_{\odot}$ towards the West limb was performed for PSP to squeeze into the Metis POS, which extended from about $6.3$ to $13.3$ R$_{\odot}$. Metis observations are supported by a $3$D MHD modeling of the solar corona developed by Predictive Science Incorporated \citep[PSI;][]{2018NatAs...2..913M}. This is based on photospheric magnetic field measurements acquired by the Helioseismic and Magnetic Imager \citep[HMI;][]{2012SoPh..275..207S} on board the Solar Dynamics Observatory \citep[SDO;][]{2012SoPh..275....3P} and driven by a heating mechanism fed by the dissipation of low-frequency Alfv\'en waves.

\begin{figure}[h]
	\begin{center}
		\includegraphics[width=\linewidth]{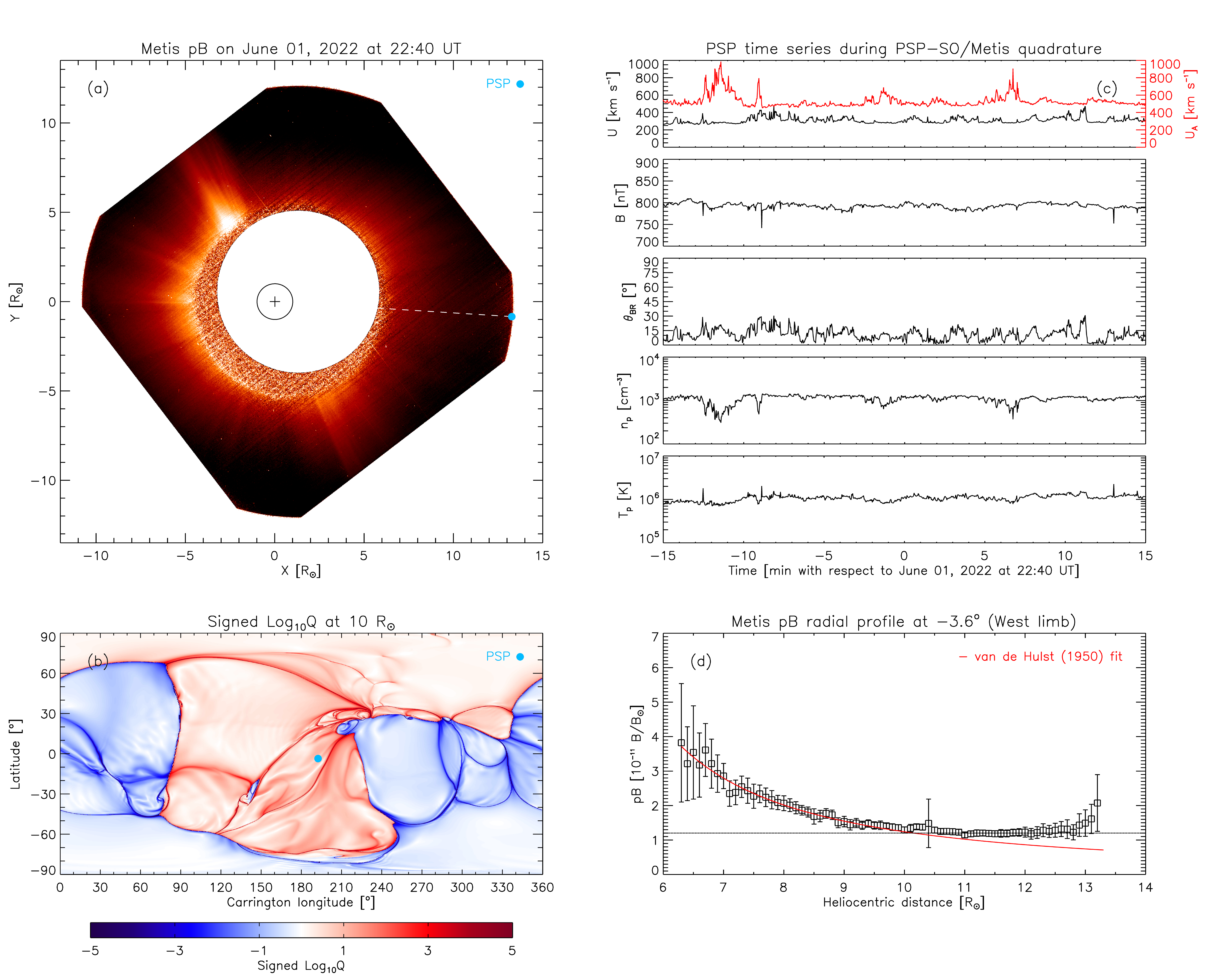}
	\end{center}
	\caption{(a) Metis pB image on June $01$, $2022$ at $22$:$40$ UT with PSP marked in its POS as a blue dot; the PSP direction is indicated by a white dashed line. (b) Projection of the PSP position (blue dot) onto the Carrington map of the squashing factor $Q$ at $10$ R$_{\odot}$. (c) PSP magnetic field and plasma time series (from top to bottom: wind bulk and Alfv\'en speed, magnetic field intensity and angle with the radial, proton density and temperature) during the quadrature with SO. (d) Radial profile of the Metis pB observations at the latitude corresponding to the PSP location (open squares), along with the \citet{1950BAN....11..135V} power-law fit (red line); the horizontal line at $1.2\times10^{-11}$ B$/$B$_{\odot}$ represents the flattening level probably due to the contamination by the F corona and instrumental effect.}
	\label{fig:observations}
\end{figure}

Figure \ref{fig:observations}(a) shows the visible light observations of the coronal polarized Brightness (pB) obtained with Metis during the quadrature with PSP, marked with a blue dot, and calibrated according to \citet{ref:calibration_vl}. PSP, $3.6^{\circ}$ below the equatorial plane, was immersed in a diffuse coronal region devoid of emission-enhanced (i.e., denser) structures. Indeed, Figure \ref{fig:observations}(b), which displays the Carrington map of the squashing factor $Q$ \citep{2011ApJ...731..111T} at $10$ R$_{\odot}$, clearly discloses that PSP sampled a low-latitude equatorial coronal hole, which, in this case, is near the outer positive polarity of a decaying active region from the previous rotation. Due to the pseudostreamer lobes bracketing the open flux (Figure \ref{fig:observations}(b)), however, the solar wind flow impinging on PSP was slow \citep{1994ApJ...437L..67W}, as shown in the top panel of Figure \ref{fig:observations}(c), which overall displays PSP plasma and magnetic field measurements acquired on a $0.5$-hour long interval around the quadrature with SO/Metis (corresponding to $\pm0.76^{\circ}$ longitude with respect to the Metis POS). Specifically, from top to bottom, are the solar wind and Alfv\'en speed, magnetic field vector strength and inclination with respect to the radial direction, and proton density and temperature. This is a case of highly field-aligned ($\theta_{BR}\sim11^{\circ}$) and sub-Alfv\'enic slow ($U\sim310$ km s$^{-1}<U_{A}$) solar wind, which is additionally quite homogenous and stationary.

Finally, Figure \ref{fig:observations}(d) shows the pB radial profile at the latitude in the PSP direction (indicated by a white dashed line in Figure \ref{fig:observations}(a)). Relying on the approach initially advanced by \citet{1950BAN....11..135V} and largely used with Metis observations \citep[e.g.,][]{2021A&A...656A..32R,2023PhPl...30b2905A}, the coronal electron density can be determined by inverting the coefficients of a power-law fit performed on the observed pB values. The combined effect of the contamination of the K corona by the F corona for distances larger than $10$ R$_{\odot}$ \citep[e.g.,][]{2019Natur.576..232H} and a possible instrumental contribution appearing as a rise in pB values at the outer edge of the detector Field Of View (FOV) can in principle be responsible for the flattening of the pB radial profile at heights above $10$ R$_{\odot}$ (around the value indicated by the horizontal line). Accordingly, the power-law fitting has been performed up to heights of $10$ R$_{\odot}$ and then extrapolated beyond that (red line in Figure \ref{fig:observations}(d)). Anticipating the results presented in the next section, it is already worth mentioning here that the electron density profile thus derived was checked and validated with the different technique by \citet{2001ApJ...548.1081H} and was verified to scale approximately as $r^{-2}$ at large distances from the Sun, at the outer edge of the Metis FOV (i.e., $r>10$ R$_{\odot}$).

In order to derive an empirical coronal flow model from the joint SO/Metis -- PSP observations of the solar corona, the basic equations of solar wind theory need to be solved. A steady, radial flow of an ideal plasma is fully described by the equations for mass, momentum and energy conservation below:

\begin{equation}
	\frac{d}{dr}(\rho UA)=0;
	\label{eq:conservation_mass}
\end{equation}
\begin{equation}
	\rho U\frac{dU}{dr}=-\frac{dp}{dr}-\rho\frac{GM_{\odot}}{r^{2}};
	\label{eq:conservation_momentum}
\end{equation}
\begin{equation}
	U\frac{dp}{dr}=-\gamma p\nabla\cdot\mathbf{U}+(\gamma-1)\rho\epsilon,
	\label{eq:conservation_energy}
\end{equation}

\noindent where $\rho$ is the mass density, $\mathbf{U}$ is the solar wind velocity, $A=f(r)r^{2}$ is the cross-sectional area of the flow tube connecting the equatorial coronal hole to the heliosphere (with $f(r)$ being the expansion factor), $p=nk_{B}T$ is the thermal pressure (with $n$, $T$, and $k_{B}$ being the number density, temperature and Boltzmann constant, respectively), $G$ is the gravitational constant, $M_{\odot}$ is the solar mass, $\gamma$ is the adiabatic index, $\epsilon$ is the heating per unit mass, and $r$ is the radial coordinate. Note that the empirical model assumes only a very hot thermal plasma at the origin and that no distributed heating occurs due to the dissipation of turbulence. The heating rate $\epsilon$ is primarily due to low-frequency nearly-incompressible turbulence and expressed as \citep{1999ApJ...523L..93M}

\begin{equation}
	\epsilon\equiv\frac{1}{2}\alpha\frac{\langle|\mathbf{z}^{+}|^{2}\rangle^{2}\langle|\mathbf{z}^{-}|^{2}\rangle+\langle|\mathbf{z}^{+}|^{2}\rangle\langle|\mathbf{z}^{-}|^{2}\rangle^{2}}{\lambda},
	\label{eq:heating_rate}
\end{equation}

\noindent where $\alpha$ is the K\'arm\'an-Taylor constant \citep[usually around $0.125$, see][and references therein]{2014ApJ...788...43U}, $(1/2)\langle|\mathbf{z}^{\pm}|^{2}\rangle$ are the energies associated with the outward/inward fluctuations given in terms of Els\"asser variables, $\mathbf{z}^{\pm}=\mathbf{u}\pm\frac{\mathbf{b}}{\sqrt{\mu\rho}}$ (with $\mathbf{u}$ and $\mathbf{b}$ being the velocity and magnetic field fluctuations around the corresponding mean fields, $\langle\mathbf{U}\rangle$ and $\langle\mathbf{B}\rangle$, respectively, and $\mu$ the magnetic permeability), $\lambda$ is the turbulence correlation length, and $\langle...\rangle$ indicates time averaging over the data sample. Assuming that most of the turbulent energy comes from Alfv\'enic fluctuations, the system will be characterized by a balance between kinetic and magnetic energies, i.e., $\langle|\mathbf{u}|^{2}\rangle\sim\langle|\mathbf{b}|^{2}\rangle$. Equation (\ref{eq:heating_rate}) can thus be conveniently rewritten as \citep{2017ApJ...851..117A}

\begin{equation}
	\epsilon=\frac{E_{b}^{3/2}}{\left[C_{k}\log\left(\frac{1}{k_{inj}\lambda_{b}}\right)\right]^{3/2}\lambda_{b}},
	\label{eq:cascade_rate}
\end{equation}

\noindent where $E_{b}=\langle|\mathbf{b}|^{2}\rangle/(\mu\rho)$ and $\lambda_{b}$ are the energy and correlation length associated with magnetic field fluctuations, respectively, $C_{k}=1.6$ is the Kolmogorov constant, and $k_{inj}=1.07\times10^{-9}$ km$^{-1}$ is the large-scale injection wavenumber corresponding to the solar rotation frequency (related to each other by the Sun's rotational velocity). Equation (\ref{eq:cascade_rate}) is derived on the basis of a dimensional analysis of the power spectral density of the turbulent magnetic energy $E_{b}$, assuming that it scales with the wavenumber $k$ as $Ak^{-1}$ and $C_{k}\epsilon^{2/3}k^{-5/3}$ (with $A$ being a constant) in the energy-containing and inertial ranges, respectively, imposing equality of the spectrum branches at the frequency break $k_{b}$ and integrating from $k_{inj}$ to $k_{b}$. The only assumptions underlying the expression in Equation (\ref{eq:cascade_rate}) are thus a Kolmogorov-like \citep{1941DoSSR..30..301K} power law at fluid scales and that the correlation length $\lambda_{b}$ corresponds to the spectral break $k_{b}$ separating injection and inertial ranges. Although customarily used \citep[e.g.,][]{2022ApJ...938L...8T}, it is worth noting that it is a rough estimate of the heating rate that does not take into account, for example, the anisotropy in the solar wind turbulence fluctuations. In addition, in the Iroshnikov-Kraichnan picture of turbulence \citep{1963AZh....40..742I,1965PhFl....8.1385K}, with a shallower spectrum scaling as $k^{-3/2}$ as usually observed closer to the Sun \citep[e.g.,][]{2020ApJS..246...53C,2020ApJS..246...55D,2021ApJ...912L..21T}, the value for $\epsilon$ as provided by Equation (\ref{eq:cascade_rate}) would be a factor of $\lambda_{b}^{-1/4}$ larger. Completing Equations (\ref{eq:conservation_mass})--(\ref{eq:conservation_energy}) is the conservation of magnetic flux, $\frac{d}{dr}(BA)=0$.

Integrating the equations for mass and magnetic flux conservation, it follows that the large-scale trends of the solar wind parameters (i.e., $U$, $\rho$, and $B$) can be estimated based on the values at some reference point $r=r_{0}$. Specifically, from (i) the estimates of the coronal mass density $\rho(r)=0.95m_{p}n_{e}$ \citep[for a fully ionized plasma with $2.5\%$ helium,][and where $m_{p}$ is the proton mass]{2020NatAs...4.1134M} as a function of the heliocentric distance derived from the pB Metis observations, (ii) the PSP plasma and magnetic field measurements at $r_{0}=13.3$ R$_{\odot}$, and (iii) the expansion factor $f(r)$ provided by the PSI $3$D MHD simulations (which results in only a weak dependence on height, i.e., $f(r)\sim1$, away from the strong fields at the base of the corona due to the rooting of the open flux near a low-lying pseudostreamer-type configuration), it is possible to empirically evaluate the solar wind speed and magnetic field strength radial profiles, $U(r)$ and $B(r)$, from $6.3$ to $13.3$ R$_{\odot}$. The Alfv\'en speed $U_{A}(r)$ and the solar wind energy flux $F_{w}(r)$ can be then immediately quantified in the Metis FOV as $U_{A}=B/\sqrt{\mu\rho}$ and $F_{w}=\frac{1}{2}\rho U^{3}$. In other words, in this joint SO/Metis -- PSP analysis, Metis provides the radial trends for all the magnetofluid and turbulence parameters, while PSP, which definitely entered the corona (indeed the Alfv\'en mach number $M_{A}=U/U_{A}\sim0.59$), sets the (absolute) value at $r_{0}$.

The energy of the magnetic field fluctuations $E_{b}$ and, in turn, the turbulent cascade (heating) rate $\epsilon$ can be empirically estimated deeper in the solar corona at the heights observed with Metis, by propagating the measurements made locally by PSP back to the Metis FOV. For simplicity, it is assumed the propagation equation for Alfv\'en waves in an expanding flow \citep{1993A&A...270..304V}, generalized to include a dissipation term $\mathcal{D}(\langle|\mathbf{z}^{+}|^{2}\rangle^{2}\langle|\mathbf{z}^{-}|^{2}\rangle,\langle|\mathbf{z}^{+}|^{2}\rangle\langle|\mathbf{z}^{-}|^{2}\rangle^{2},\lambda)$ and possible in-situ sources of turbulence $\mathcal{S}$ \citep{1990JGR....9510291Z,1996JGR...10117093Z,2022ApJ...928..176W}

\begin{equation}
	\frac{\partial\mathbf{z}^{\pm}}{\partial t}+(\mathbf{U}\pm\mathbf{U_{A}})\cdot\nabla\mathbf{z}^{\pm}+\mathbf{z}^{\mp}\cdot\nabla(\mathbf{U}\mp\mathbf{U_{A}})+\frac{1}{2}(\mathbf{z}^{-}-\mathbf{z}^{+})\nabla\cdot(\mathbf{U_{A}}\mp\frac{1}{2}\mathbf{U})=-\mathcal{D}(-)+\mathcal{S}.
	\label{eq:wave_propagation}
\end{equation}

\noindent If the further simplifying assumption of linearity and a slowly varying background is made, Equation (\ref{eq:wave_propagation}) can be solved in the WKB approximation to retain an analytically tractable solution. In the presence of a nonuniform but stationary flow, the wave action flux $S=\frac{1}{2}\rho|\mathbf{z}^{+}|^{2}/\omega$, namely the wave energy flux per unit frequency $\omega$, is conserved \citep[see, e.g.,][and references therein]{1993A&A...270..304V}. Assuming an outwardly directed magnetic field, wave action conservation in the stationary WKB limit translates into

\begin{equation}
	\nabla\cdot(\mathbf{U}+\mathbf{U_{A}})S=0.
	\label{eq:wave_action_flux_conservation}
\end{equation}

\noindent Considering that the wave frequency in the absolute (stationary) frame, that is the wave eigenfrequency $\omega_{0}$, is an invariant, the frequency in the plasma frame $\omega$ is given by

\begin{equation}
	\omega=\frac{U_{A}}{U+U_{A}}\omega_{0}.
	\label{eq:doppler_effect}
\end{equation}

\noindent Bearing furthermore in mind the mass flux conservation $\rho Ur^{2}=const$ and the divergence formula in spherical coordinates for a purely radial vector $\nabla\cdot\mathbf{F}=\frac{1}{r^{2}}\frac{\partial}{\partial r}(r^{2}F_{r})$, Equation (\ref{eq:wave_action_flux_conservation}) simplifies to

\begin{equation}
	\frac{(U+U_{A})^{2}}{UU_{A}}|\mathbf{z}^{+}|^{2}=C,
	\label{eq:wave_energy_conservation}
\end{equation}

\noindent with $C$ a constant. Equation (\ref{eq:wave_energy_conservation}) represents, in the WKB limit, a very useful expression for the radial dependence of the Alfv\'en wave energy, once estimated at a reference point $r=r_{0}$ (i.e., at PSP). It predicts that the outwardly propagating Alfv\'en mode energy peaks (and, therefore, turbulence is maximum) where $U=U_{A}$, i.e., at the Alfv\'en point. In the case of pure Alfv\'en waves \citep[i.e., $\sigma_{c}=\pm1$ and $\sigma_{r}=0$, where $\sigma_{c}$ and $\sigma_{r}$ are the normalized cross-helicity and residual energy, which measure the imbalance between outward and inward modes and between kinetic and magnetic energies, respectively, e.g.,][]{1995SSRv...73....1T}, the above equation can also be used for the energy of the magnetic fluctuations $E_{b}$. Associating, to a first approximation, the wavelength of turbulence with the expansion of the flow tube, i.e., $\lambda_{b}\propto\sqrt{A}\sim r$, a complete set of functional forms for estimating the energy transfer rate at the coronal heights observed with Metis is provided.

In the PSP $0.5$-hour interval considered in the present analysis (Figure \ref{fig:observations}(c)), $\sigma_{c}=0.97$ and $\sigma_{r}=-0.10$, suggesting the overwhelming presence of Alfv\'en waves in the downwind direction. Furthermore, the power spectrum of the magnetic field fluctuations may be compatible with the $5/3$ Kolmogorov turbulence (the power-law fit performed at fluid scales to $E_{b}$ in fact returns a spectral index of $1.61\pm0.04$). It follows that Equation (\ref{eq:cascade_rate}) can be safely back-projected from the PSP location, which is below the Alfv\'en point ($U<U_{A}$, see Figure \ref{fig:observations}(c)), into the Metis FOV and thus the coronal heating rate directly derived in the solar corona.

\section{Empirical slow solar wind and WKB-like turbulence model}
\label{sec:empirical_model}
\begin{figure}[h]
	\begin{center}
		\includegraphics[width=\linewidth]{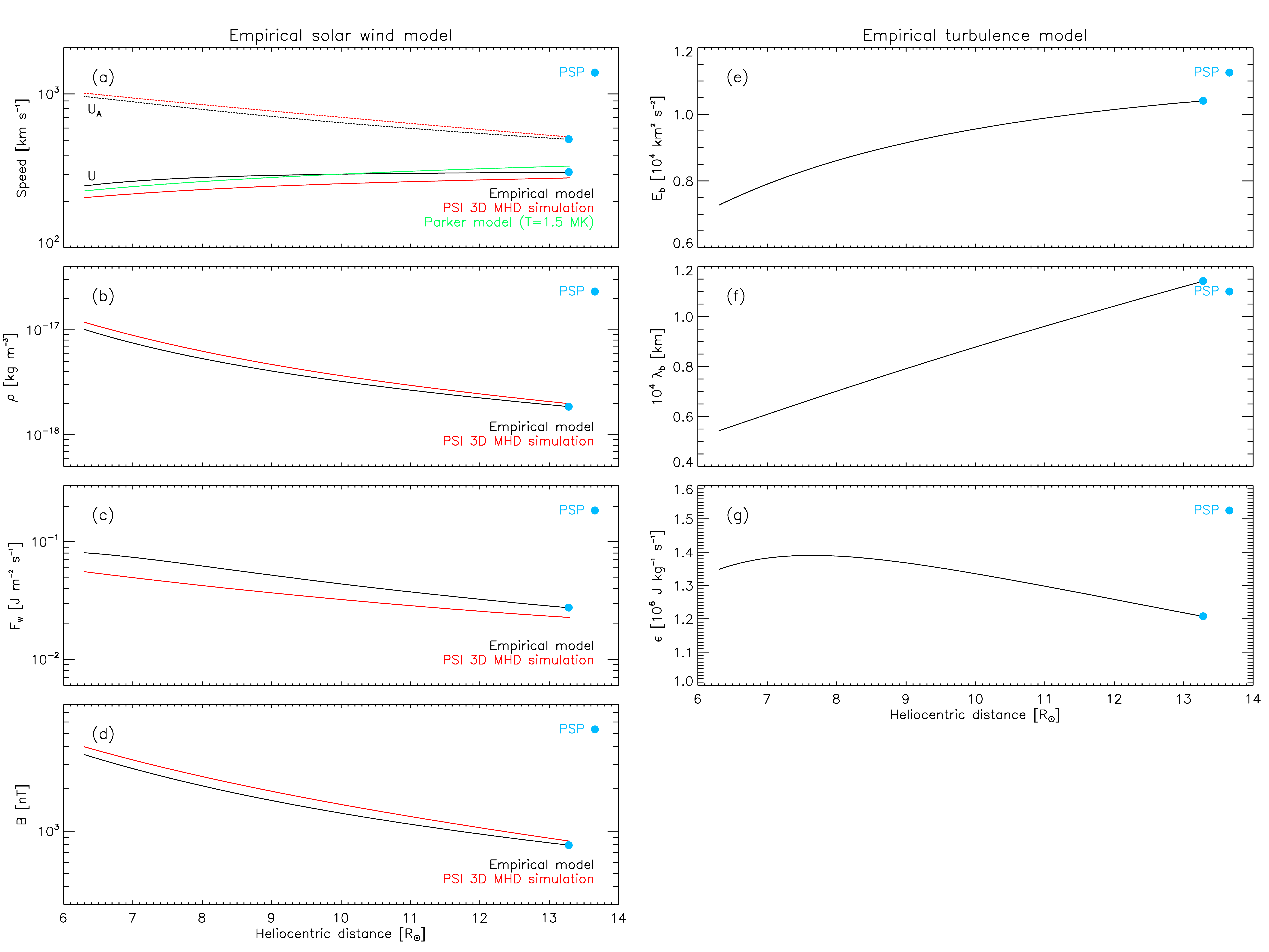}
	\end{center}
	\caption{Data-based (black) and physics-based (red) modeling of the radial evolution of solar wind $U$ (thick lines) and Alfv\'en $U_{A}$ (thin lines) speed (a), plasma mass density $\rho$ (b), solar wind flux energy $F_{w}$ (c), magnetic field $B$ (d), turbulent energy $E_{b}$ (e) and correlation length $\lambda_{b}$ (f), and heating rate $\epsilon$ (g) in the slow coronal flow jointly observed with SO/Metis and PSP during their quadrature on June $01$, $2022$. The blue dot in each panel refers to the values obtained with PSP measurements. The Parker classical model for an expanding isothermal $1.5$ MK corona is shown as a green line in (a).}
	\label{fig:wind_turbulence_model}
\end{figure}

Figure \ref{fig:wind_turbulence_model} shows the solar wind (left panels) and turbulence (right panels) models empirically derived according to the methodological approach described in the previous section (PSP measurements-based estimates are indicated by a blue dot). Specifically, Figures \ref{fig:wind_turbulence_model}(a)--(d) display the modeled coronal plasma flow speed, mass density and energy flux, and coronal magnetic field, respectively, compared with the corresponding quantities obtained from the PSI $3$D MHD simulations (red lines). In Figure \ref{fig:wind_turbulence_model}(a), also depicted are the Alfv\'en speed (thin lines) and, for comparison, the spherically symmetric expansion of an isothermal $T=1.5\times10^{6}$ K corona \citep[][green line]{1958ApJ...128..664P}.

Immediately evident is the good agreement, both in radial trends and absolute values, between the empirical model and the outcome of the MHD simulations. This allows a first, prompt consideration. In fact, the two models, in addition to being independently derived, are also the result of somewhat opposite approaches. The MHD simulation is based on a forward modeling driven by (remote) photospheric measurements and the dissipation of Alfv\'enic turbulence as the mechanism for heating the coronal plasma. The empirical model obtained in this work, on the other hand, is a backward extrapolation based on PSP in-situ measurements and the solution of the Euler equations in the WKB limit. Thus, the good agreement between the two models (for both fluid and magnetic parameters) indicates, on the one hand, the accuracy of the present analysis and especially of the Metis observations and, on the other hand, that the aforementioned assumed physical mechanism underlying the heating and subsequent acceleration of the coronal wind complies with observations, again identifying it as the most likely mechanism of coronal heating (in contrast to the alternative high-frequency resonant dissipative processes). More quantitatively, the empirically-derived and simulation-based models differ (on average) by $16\%$ for the flow speed, $12\%$ for the plasma density and magnetic field, and $37\%$ for the wind energy flux (for which, as a derived quantity, the discrepancies associated with the fundamental parameters $U$ and $\rho$ obviously widen). In discussing in greater detail the expansion rate of the wind, Figure \ref{fig:wind_turbulence_model}(a) clearly shows that, although at the distances observed with Metis the coronal plasma is already largely accelerated, some residual acceleration still persists: in fact, the wind speed increases from $\sim250$ to $\sim310$ km s$^{-1}$ over a range of $7$ R$_{\odot}$. Nevertheless, as evidenced by the basically good agreement with Parker's model, this acceleration may just be thermally driven, which indicates that collisionless field-particle interaction mechanisms for heating (and thereby boosting) the wind are most effective lower than the distances observed with Metis \citep[as, for instance, shown in][]{2023A&A...670L..18T}. Rough estimates of the power-law radial dependence of the physical quantities displayed in Figures \ref{fig:wind_turbulence_model}(a)--(d), which follow from the continuity equations above and the experimental finding that $\rho\sim r^{-2.21}$, can be given. It turns out that $U\sim r^{0.23}$, $F_{w}\sim r^{-1.52}$, and $B\sim r^{-1.98}$ for $6.3$ R$_{\odot}$ $<r<13.3$ R$_{\odot}$.

Figures \ref{fig:wind_turbulence_model}(e)--(g) show the empirically modeled turbulence energy, correlation length and cascade rate, respectively. As a consequence of Equation (\ref{eq:wave_energy_conservation}) and having assumed that $\lambda_{b}\propto\sqrt{A}$, the fluctuating magnetic energy and the corresponding correlation length decrease from the PSP position towards the Sun (according to $E_{b}\sim r^{0.45}$ and $\lambda_{b}\sim r^{0.99}$, respectively). The resulting heating rate in the slow coronal plasma jointly observed with SO/Metis -- PSP keeps a relatively constant value throughout the considered heliocentric distances ($\epsilon\sim r^{-0.18}$), in agreement with earlier studies of wind equations and turbulence models \citep[e.g.,][]{2002ApJ...575..571D}, which show that the energy per unit mass does not fall off rapidly with altitude. In this regard, it is worth noting that the slight peak in $\epsilon$ observed at about $7-8$ R$_{\odot}$ is merely due to the combination of the $E_{b}$ and $\lambda_{b}$ radial trends and, hence, does not indicate any particular underlying physical process. In fact, the heating rate is expected to peak much closer in, at the temperature maximum and the sonic point (around around $2-4$ R$_{\odot}$), i.e., where plasma heating and acceleration are most effective. The average (turbulent) heating rate per unit volume $\overline{\epsilon\rho}=5.5\times10^{-12}$ J m$^{-3}$ s$^{-1}$ is somewhat lower that the estimates of $10^{-9}-10^{-11}$ J m$^{-3}$ s$^{-1}$ recently reported by \citet{2009ApJ...707.1668C} and \citet{2021ApJ...914..137S} in approximately the same altitude region ($5-45$ R$_{\odot}$). This is not surprising, since these authors provided upper limits for $\epsilon$, whereas those presented here are lower estimates. In fact, the WKB approximation does not take into account the undoubted dissipation of $\mathbf{z}^{+}$ modes. As a result, the rate at which their energy increases with distance has to be lower than the WKB prediction. Hence, from the PSP position down into the solar corona, this means that their energy $E_{b}$ would fall off less rapidly, resulting in an energy transfer rate higher than estimated here in the WKB limit. Moreover, under the likely scenario that deeper into the corona the spectrum of the magnetic field fluctuations more closely resembles the turbulence \`a la Iroshnikov-Kraichnan, for the above, the energy transfer rate would be larger by a factor of $\lambda_{b}^{-1/4}$, resulting in an average rate per unit volume of $5.2\times10^{-11}$ J m$^{-3}$ s$^{-1}$, thus at the lower bound of the range of values previously reported in the literature.

The coronal energy loss $H$, expressed as a flux, can be estimated by integrating the heating rate in the Metis FOV, i.e., from $6.3$ to $13.3$ R$_{\odot}$,

\begin{equation}
	H=\int\epsilon(r)\rho(r)dr=2.7\times10^{-2}\,\mathrm{J}\,\mathrm{m}^{-2}\,\mathrm{s}^{-1}.
	\label{eq:energy_loss}
\end{equation}

\noindent It turns out that $H$ is $33\%$ of the total energy flux of the solar wind ($F_{w}=8.1\times10^{-2}$ J m$^{-2}$ s$^{-1}$). This is expected, since, as mentioned just above, most of the acceleration has already taken place at the heights observed with Metis. This energy deposition feeds the observed residual acceleration of the slow solar wind and heats the non-adiabatic coronal plasma. Indeed, in a collisionless plasma with no additional heating, based on the conservation of the Chew-Goldberger-Low \citep[CGL;][]{1956RSPSA.236..112C} invariants, the total temperature should scale as $r^{-4/3}$ with distance. This is not observed either in the corona or in interplanetary space \citep[where a radial scaling greater than that predicted by the adiabatic theory of plasma thermal dynamics is found, e.g.,][]{2019ApJ...879...32Z}, implying that plasma heating (via turbulence dissipation) occurs, although less significantly even at distances far beyond where the heating and acceleration processes are most at work (around the sonic point). It is well known that the deficit in internal energy implied by an adiabatic expansion persists at least to $1$ au, and in fact well beyond \citep{1995GeoRL..22..325R,1999PhRvL..82.3444M}.

As a final remark, although the present WKB-like analysis is important as it provides a model-independent and empirical lower limit of the coronal heating rate (thanks to the coordinated Metis/SO -- PSP observations), it should be recalled that it is well known that the radial evolution of the turbulent solar wind can resemble WKB theory even when driven by waves \citep{1989JGR....94.6899R} or when turbulent dissipation and shear driving is included \citep{1996JGR...10117093Z}. It is also worth noting that the WKB approximation does not hold in the expanding solar corona. In fact, the wave-only WKB theory predicts no dissipation, whereas turbulence has largely been dissipated far below to accelerate the wind to supersonic speeds. It is therefore evident that a more realistic description of the turbulence evolution from the subsonic wind to the PSP location is needed to compare directly with the Metis observations and thus accurately estimate the (turbulent) heating rate in the solar corona. A variety of turbulence models \citep[e.g.,][]{2010ApJ...708L.116V,2011JGRA..116.8105O} may be useful as extension of the present simplified WKB-like model in order to explain observations when linked self-consistently with large scale MHD equations.

\acknowledgments
Solar Orbiter is a space mission of international collaboration between ESA and NASA, operated by ESA. D.T. was partially supported by the Italian Space Agency (ASI) under contract 2018-30-HH.0. G.P.Z., L.A., and L.-L.Z. acknowledge the partial support of a NASA Parker Solar Probe contract SV4-84017, an NSF EPSCoR RII-Track-1 Cooperative Agreement OIA-2148653, and a NASA IMAP grant through SUB000313/80GSFC19C0027. S.B and J.K. acknowledge support from NASA contract NNN06AA01C. The Metis program is supported by ASI under contracts to the National Institute for Astrophysics and industrial partners. Metis was built with hardware contributions from Germany (Bundesministerium f\"ur Wirtschaft und Energie through the Deutsches Zentrum f\"ur Luft- und Raumfahrt e.V.), the Czech Republic (PRODEX) and ESA. The Metis team thanks the former PI, Ester Antonucci, for leading the development of Metis until the final delivery to ESA. D.T. also wishes to thank her for the helpful comments to the paper. The Metis data analyzed in this paper are available from the PI on request. Parker Solar Probe data was downloaded from the NASA's Space Physics Data Facility (\href{https://spdf.gsfc.nasa.gov}{https://spdf.gsfc.nasa.gov}).
\par



\end{document}